\documentclass[preprintnumbers,amsmath,amssymb,aps]{revtex4}
\usepackage{graphicx}

\begin{document}

\preprint{Tl dots}

\title{Self-trapping nature of Tl nanoclusters on Si(111)-7$\times$7 surface}
\author{C. G. Hwang}
\author{N. D. Kim}
\author{G. Lee}
\author{S. Y. Shin}
\author{J. S. Kim}
\author{J. W. Chung}\email{jwc@postech.ac.kr}
\address{Physics Department, Pohang University of Science and Technology, Pohang 790-784, Korea}

\date{\today}

\begin{abstract}
We have investigated electronic and structural properties of thallium (Tl) nanoclusters formed on
the Si(111)-7$\times$7 surface at room temperature (RT) by utilizing photoemission spectroscopy
(PES) and high-resolution electron-energy-loss spectroscopy (HREELS) combined with first principles
calculations. Our PES data show that the state S2 stemming from Si restatoms remains quite inert
with Tl coverage $\theta$ while S1 from Si adatoms gradually changes, in sharp contrast with the
rapidly decaying states of Na or Li nanoclusters. No Tl-induced surface state is observed until
$\theta$=0.21 ML where Tl nanoclusters completely cover the faulted half unit cells (FHUCs) of the
surface. These spectral behaviors of surface states and a unique loss peak L$_2$ associated with Tl
in HREELS spectra indicate no strong Si-Tl bonding and are well understood in terms of gradual
filling of Si dangling bonds with increasing $\theta$. Our calculational results further reveal
that there are several metastable atomic structures for Tl nanoclusters at RT transforming from
each other faster than 10$^{10}$ flippings per second. We thus conclude that the highly mobile Tl
atoms form self-trapped nanoclusters within FHUC at RT with several metastable phases. The mobile
and multi-phased nature of Tl nanoclusters not only account for all the existing experimental
observations including the fuzzy scanning tunneling microscope images and a dynamical model
proposed by recent x-ray study but also provides an example of self-trapping of atoms in a
nanometer-scale region.
\end{abstract}


\maketitle

\section{INTRODUCTION}
Since the success of atom-trapping in microwave cavity using the field from a single photon,\cite{Haroche} efforts have
been continued to cool-down atoms with a better spatial resolution motivated by numerous applications including
Bose-Einstein condensates, high-precision atomic clocks, and scalable quantum computers.\cite{Vul,Hood} Despite such
elaborate research endeavor, the confinement of a single atom has been limited only to a few tens of micrometer. We,
however, report that nature allows the trapping of several thallium (Tl) atoms into a region of nanometer scale in the
form of a nanocluster on the Si(111)-7$\times$7 surface at room temperature (RT). The trapping of Tl atoms appears to
be distinctly different from typical features of Tl$-$Si atomic bonding revealing a remarkably mobile character of Tl
nanoclusters and significant inertness on substrate surface electronic states.

\begin{figure}[b]
\includegraphics{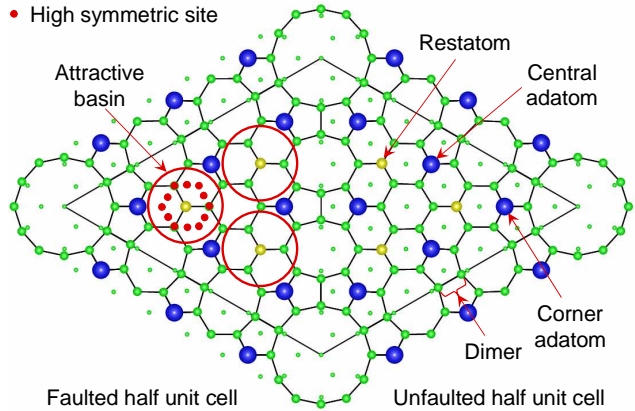}
\caption{\label{fig:basin}(Color online) Structure of the Si(111)-7$\times$7 unit cell. Several high symmetric sites
(red dots) around restatoms within the attractive basins (red circles) are possible binding sites of adsorbates. (Ref.
\onlinecite{CK})}
\end{figure}

Because of its unique atomic arrangement of the unit cell consisting of a faulted half unit cell (FHUC) and an
unfaulted half unit cell (UFHC) as depicted in Fig.~\ref{fig:basin}, the Si(111)-7$\times$7 surface has been used as a
fascinating template to fabricate a crystalline array of self-assembled nanoclusters of various atomic species. Most
adsorbates upon forming nanoclusters are found to occupy high symmetric sites around Si restatoms known as attractive
basins (red circles in Fig.~\ref{fig:basin}).\cite{CK} Those nanoclusters formed on this surface at RT exhibit
atomically well-resolved scanning tunneling microscopy (STM) images with six atoms in one or both half unit cells and
appear to be semiconducting.\cite{Li,Wu,Li2} Tl nanoclusters, however, has been known to be exceptional to this trend
showing fuzzy STM images and reported to be metallic.\cite{Vitali} It has been proposed that each Tl nanocluster
contains nine mobile Tl atoms, instead of six as for most other nanoclusters, formed only in FHUC to account for such
distinct features of Tl nanoclusters.\cite{Vitali,Zotov} Moreover studies on electronic band structures of most
nanoclusters other than Tl suggest a strong chemical bonding between adsorbates and Si atoms accompanying significant
displacement of central Si atoms by adsorbates.\cite{Zhang,Ahn,Wei} Such a strong chemical bonding may explain why the
STM images of nanoclustares are so well resolved atomically for those nanoclusters except Tl nanoclusters.

Motivated by such peculiar features of Tl nanoclusters, we have investigated electronic and structural properties of Tl
nanoclusters formed on the Si(111)-7$\times$7 surface by utilizing photoemission spectroscopy (PES) and high resolution
electron energy loss spectroscopy (HREELS). We also have carried out first principle total energy calculation to
estimate thermal stability of atomic arrangement of Tl nanoclusters at RT. We find that the substrate surface states,
especially associated with Si restatoms, are quite inert to Tl adsorption until the completion of formation of
nanoclusters in contrast to nanoclusters of other atomic species. Such a remarkable inertness of surface states
together with invisible semiconducting band gap in HREELS spectra suggest a distinctly weak chemical bonding between Tl
and Si atoms. This also may eliminate a possibility of displacing Si center adatom by Tl, which is often observed for
most other nanoclusters formed on Si(111)-7$\times$7 surface. Our results of first principle total energy calculation
for several plausible atomic models of Tl nanoclusters, in fact, support such a scenario by revealing relatively small
energy differences between different atomic models, and thus a significant hopping rate between different atomic
configurations within the same attractive basin at RT. We find a "dynamic trapping state" of Tl nanoclusters in a sense
that a single Tl nanocluster consisting of nine Tl atoms is rapidly changing its atomic configuration due to the low
diffusion barriers between neighboring high symmetry sites in the attractive basin while residing only in a FHUC. We
present experimental evidence and discuss physical implication of the dynamic trapping state.

\section{EXPERIMENTAL DETAILS}
The PES chamber used to measure the valence band of our sample has a high intensity He I discharge
lamp (Omicron HIS-13 ) and a SPECS Phoibos-100 electron analyzer with an optimal resolution of 110
meV at RT. The HREELS system utilizes a Leybold-Heraeus ELS-22 spectrometer with an optimum
resolution of 19 meV. Both the PES and the HREELS chambers equipped with several surface diagnostic
probes including low energy electron diffraction (LEED) have been maintained with a base pressure
of less than 1$\times$10$^{-10}$ Torr during the entire course of measurements. We have prepared
our sample by using a high-doped n-type Si(111) wafer with a resistivity of 2 $\Omega$cm. A Tl
source was made by wrapping a small piece of Tl with a tungsten wire. We have thoroughly degassed
sample and Tl source with the chamber pressure controlled under 3$\times$10$^{\text{-}10}$ Torr for
several hours. The clean Si(111)-7$\times$7 phase was obtained after cleaning the sample with a
well known recipe, i.e., heating up to 1200 $^{\circ}$C for 10 seconds followed by annealing at 800
$^{\circ}$C for about 5 minutes. The cleaned sample showing a well defined 7$\times$7 LEED pattern
was then cooled to RT before depositing Tl. We have first monitored intensity variation of
fractional-order LEED spots during the continuous deposition of Tl atoms onto the clean
Si(111)-7$\times$7 surface at RT. The seventh order LEED spots becomes sharper and stronger as Tl
coverage $\theta$ increases and reaches a maximum in intensity at $\theta$=0.2 monolayers (MLs) due
to the constructive interference from a lattice-like array of Tl nanlclusters. For $\theta\geq$0.2
ML, the spots become weaker and fuzzier with increasing coverage indicating an increasing disorder
due to the extra Tl atoms added on the Tl nanoclusters.\cite{Hwang} Thus the completion of
formation of a crystalline array of Tl nanoclusters on the Si(111)-7$\times$7 surface were
relatively easily detected by the local maximum in intensity of the (1/7 0) LEED spot at
$\theta_c$=0.2 ML in agreement with previous STM study.\cite{Vitali}

\begin{figure}[t]
\includegraphics{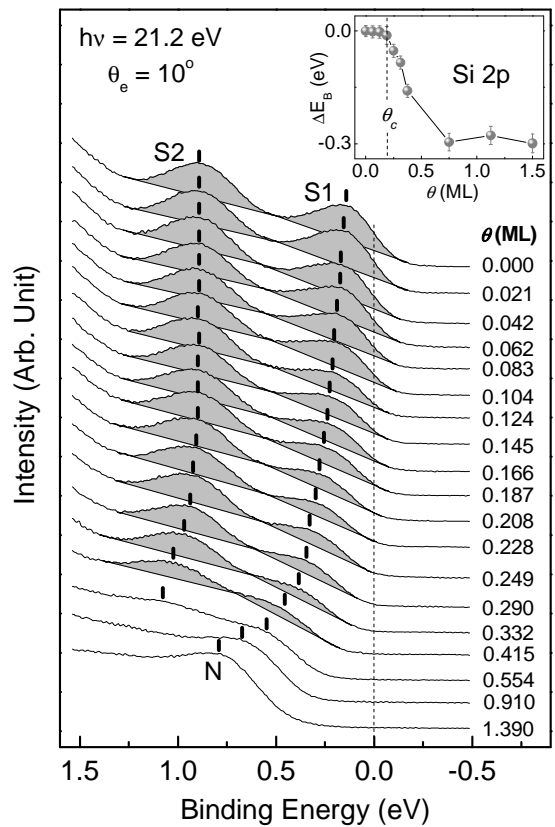}
\caption{\label{fig:pes}Progressive change of surface states S1 and S2 in photoemission spectra of
the valence band as a function of Tl coverage $\theta$. The spectra were taken with a photon energy
of 21.2 eV at an emission angle of 10$^{\circ}$ and their binding energy were corrected by Si 2$p$
core-level shift shown in inset. The surface states appear to be rather inert to Tl adsorption
until at $\theta_c$=0.2 ML where Tl nanoclusters are spread uniformly throughout the surface. A new
surface state (denoted as N) begins to show up at $\theta\geq$0.91 ML. }
\end{figure}

\section{RESULT AND DISCUSSION}
In Figure~\ref{fig:pes}, we present a progressive spectral change of the valence band of the
Si(111)-7$\times$7 surface at RT with increasing $\theta$. The spectra were measured at an emission
angle of 10$^{\circ}$ from the surface normal. One notes the well-defined surface states S1 and S2
from the clean Si(111)-7$\times$7 surface at binding energies of 0.15 and 0.87 eV below the Fermi
level. These states has been associated with central Si adatoms and Si restatoms,
respectively.\cite{Uhr} The contribution from corner Si adatoms has been reported to appear with
binding energy of 0.5 eV.\cite{Uhr} This contribution apparently is not well resolved at RT and
causes the line shape of S1 asymmetric. The binding energy of the spectra in Fig.~\ref{fig:pes} has
been corrected by taking the band bending effect into account from the Si 2$p$ core-level shift as
a function of $\theta$ (see inset). We find that the bulk component of the Si 2$p$ core-level
remains unaffected by Tl adsorption until $\theta_c$=0.2 ML where the array of Tl nanoclusters is
best developed, and then begins to change rapidly with $\theta$ exhibiting a band bending effect
due to the different chemical morphology from that of Tl nanoclusters.

We notice in Fig.~\ref{fig:pes} that the two surface states are not significantly affected by the Tl adsorption at
initial stage for $\theta\leq$0.21 ML. This is especially true for S2, which is in sharp contrast with adsorption of
other metal atoms, Li and Na for example, where S2 disappears almost completely at a Na coverage as small as 0.03
ML.\cite{Ahn,Wei} We also note that no new state associated with Tl appears until $\theta$=0.91 ML where a new state N
begins to show up. We remind that a new state appears for the Na adsorption as early as at 0.03 ML.\cite{Ahn,Wei} It is
interesting to find, however, that adsorption of K shows the spectral change with K coverage quite similar to the one
caused by Tl.\cite{K05} In order to examine the spectral changes more quantitatively, we have fitted the spectra with
Gaussian peaks after subtracting the background with a polynomial function in the form of $f(x) = a(x-x_0) + b +
c(x-x_0)^{-1} + d(x-x_0)^{-2} + e(x-x_0)^{-3}$.

\begin{figure}[t]
\includegraphics{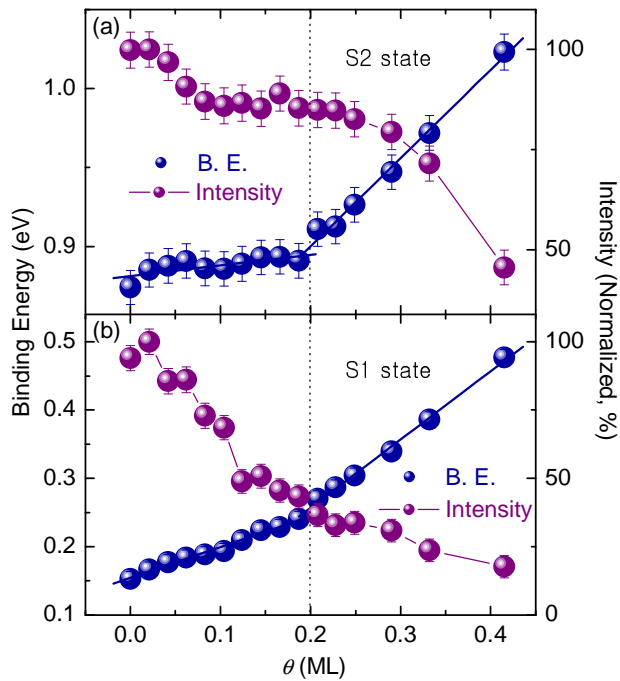}
\caption{\label{fig:analysis}(Color online) Changes of binding energy (green dots) and intensity
(red dots) of surafe state S2 (a) and S1 (b) extracted from PES data in Fig.~\ref{fig:pes} by
fitting the spectra with Gaussian peaks after subtracting a polynomial background. The changes
appear to be significantly different before and after $\theta_c$=0.2 ML.}
\end{figure}

We present our fit-results in Fig.~\ref{fig:analysis} for the states S2 in (a) and S1 in (b). Both
states reveal remarkable spectral changes as coverage crosses $\theta_c$=0.2 ML as also seen from
the structural changes in previous STM study.\cite{Vitali} While the binding energy and the
intensity of S2 changes gradually up to $\theta_c$, they change more significantly for
$\theta\geq\theta_c$. Such a trend seems to be reversed for S1. More specifically while the binding
energy of S2 (S1) is increased by 20 (90) meV at $\theta_c$, intensity of S2 (S1) is decreased by
14 (50) \%.

Now we think of several plausible mechanisms to explain such spectral changes in
Fig.~\ref{fig:analysis}. One may first consider the filling of the half-filled Si dangling bonds by
the charge transfer from Tl atoms. Here we assume no displacement of central Si adatoms by Tl since
there is no new surface state appearing at least until $\theta_c$=0.2 ML. The half-filled S1 state
pinning the Fermi level at 0.63 eV above from valence band maximum\cite{Himp} becomes increasingly
filled by the electrons from Tl atoms so as to make the state shift towards the higher binding
energy side as observed in Fig.~\ref{fig:analysis}. Similar shifts have been reported also for the
electron doped bulk crystal or K adsorbed graphene.\cite{King,Ohta} The decreased intensity of S1
by 50 \% suggests the filling of the dangling bonds of central Si adatoms only in FHUC, which is
consistent with earlier STM observation of Tl nanoclusters formed only in FHUC.\cite{Vitali} While
S1 is affected significantly by the filling, S2 state may remain inert as observed when the charge
transfer is limited mainly to the central Si adatoms.

Another possibility to explain binding energy shifts of the surface states, especially for the
remarkable binding energy shift of S1 state, is the effect driven by the enhanced inter-clusters
interaction due to the reduced separation between neighboring nanoclusters with increasing
$\theta$. El-Moghraby et. al, showed in their calculation that the reduced inter-clusters
separation results in lowering the ground level energy due to the enhanced inter-clusters
interaction.\cite{Mogh} Similar trend has been reported also for the excitonic energies of quantum
dots.\cite{Zunger} Since exciton is associated with electron-hole pair excitations, the enhanced
excitonic energy indicates the shift of valence band so that the binding energy shift of the
surface state S1 may be caused partly by the enhanced inter-clusters interaction with increasing
$\theta$. Unfortunately we can not quantify the amounts of shifts due to this interaction at
present since no relevant calculations available. We reported that the constant work function
change observed during the coverage range while the Tl nanoclusters were formed might be understood
by the enhanced inter-clusters interaction.\cite{Hwang} Although one may still think of a
possibility that a new surface state may exist quite near to the S1 and/or S2 states but is hidden
because of its weak intensity. Since the line-widths of both surface states remain almost unchanged
with $\theta$ up to 0.21 ML, however, we ruled out this possibility. The only visible new state
associated with Tl adsorption appears for $\theta\geq$0.91 ML as shown in Fig.~\ref{fig:pes}.

\begin{figure}[t]
\includegraphics{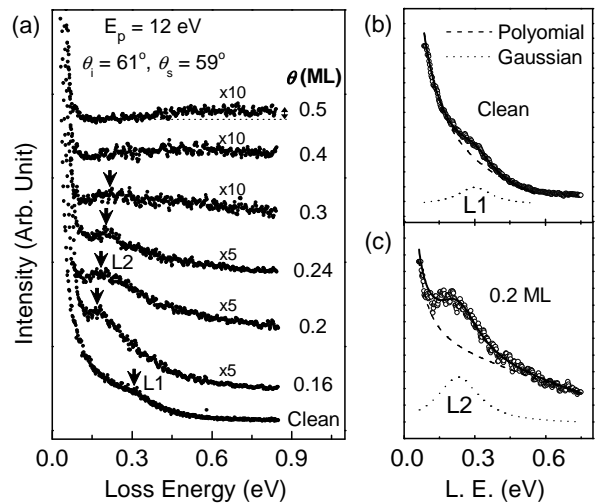}
\caption{\label{fig:eels}(a) Progressive change of HREELS spectra with Tl coverage $\theta$. The spectra were obtained
with a primary electron energy E$_p$=12 eV at 2$^{\circ}$ off from the specular direction. A unique loss peak L$_1$ (b)
of the clean Si(111)-7$\times$7 surface and a Tl-induced L$_2$ (c) are analyzed by fitting the spectra with Gaussian
peaks. One notices that L$_2$ shifts its loss energy towards the higher energy with increasing $\theta$ while its
intensity diminishes. No band gap is seen even after L$_2$ is qunched completely at coverage 0.5 ML.}
\end{figure}

Our HREELS data presented in Fig.~\ref{fig:eels} shows the spectral change with Tl adsorption. we
have fitted the HREELS spectra for a quantitative analysis as done earlier.\cite{Lait,Ahn00} The
spectral behavior turns out to be quite consistent with the explanation of our PES data based on
the filling of dangling bonds of central Si adadtoms. One first notices a Drude tail indicating the
metallic nature of the clean Si(111)-7$\times$7 surface. In Fig. \ref{fig:eels} (b), we also find a
loss peak L1 of loss energy E=0.30 eV observed only at off specular angle. Since the relatively
broad line-width ($\sim$187 meV) of L1 and small electron concentration of a Tl nanocluster, we
safely rule out possibilities of a local vibrational origin and a plasmon. It certainly can not be
a phonon since the energy is higher than the highest optical phonon energy ($\sim$57 meV) and no
multiple phonon peaks are observed. We thus attribute L1 to an interband transition between states
below and above the Fermi-level of the Si(111)-7$\times$7 surface.

The interesting feature we emphasize in Fig. \ref{fig:eels} is the presence of a loss peak L2,
which shifts toward the higher loss energy while concomitantly losing its intensity. It disappears
completely as shown by the featureless spectrum at $\theta$=0.5 ML when Tl nanoclusters no longer
maintain their unique atomic arrangement due to the extra Tl atoms. Such spectral behavior of L2
may be easily understood considering the filling of dangling bonds of Si adatoms as for the binding
energy shift of S1 state in our PES spectra in Fig.~\ref{fig:pes}. As the dangling bonds in FHUC
are gradually filled by the charge donation from Tl atoms with increasing $\theta$, the spectral
intensity or density of states (DOS) decreases accordingly in the vicinity of Fermi level.
Therefore the metallicity of the surface becomes gradually deteriorated by losing its dangling
bonds due to Tl adsorption. Since the gradual loss of the metallicity causes the weight center of
the S1 state to shift towards the higher binding side as shown in Fig.~\ref{fig:analysis}, such a
change in DOS near Fermi level should show up as a loss peak L2 shifting towards the higher loss
energy with diminishing intensity as $\theta$ increases. For $\theta\gg\theta_c$, the DOS at the
Fermi level becomes a noise level with S1 significantly quenched, which is consistent with the
HREELS spectrum at $\theta$=0.5 ML where L2 disappeared completely. Since the Tl nanoclusters are
formed only in FHUC, the surface does not show a band gap due to the remaining dangling bonds in
UHUC even though it is completely covered with Tl nanoclusters in FHUC at 0.20 ML.


In order to confirm our explanation for the experimental observations discussed above, we have carried out first
principles density-function calculations using {\it ab initio} plane wave pseudo-potential method (VASP code) in
conjunction with projector augmented wave potentials within the generalized-gradient approximations
(GGA).\cite{vasp,gga} The calculation employs a plane wave basis set with an energy cutoff of 250 eV for single $k$
point in the Brillouin zone. The unit cell in slab model consists of 1$\sim$9 Tl and 12 Si adatoms in addition to 6
layers of Si and 49 H atoms passivating Si dangling bonds in the bottom layer.

\begin{figure}[t]
\includegraphics{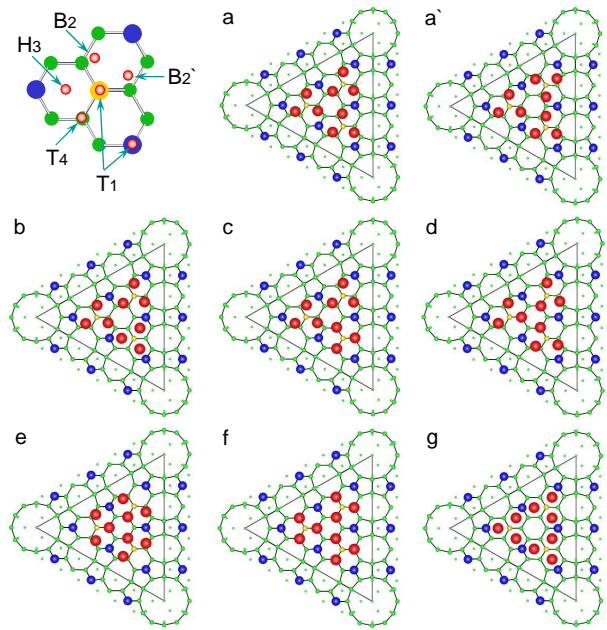}
\caption{\label{fig:model}(Color online) Total energy of eight different plausible atomic configurations (from a to g)
of Tl nanoclusters formed in FHUC have been calculated to estimate thermal stability of Tl nanoclusters at RT. Several
high symmetric sites ($B_2$, $B'_2$: bridge, $H_3$: hollow, $T_4$: on-top, and $T_1$: atop) around a restatom are
denoted as red circles in inset. Tl atoms, Si restatoms, and Si adatom in the first layer are denoted by red, yellow,
and blue circles, respectively while Si atoms in the second layers are denoted by green circles.}
\end{figure}

As a first step to find the most stable atomic structure of a Tl nanocluster, we have calculated adsorption energy for
several high symmetry sites in the attractive basin (see inset in Fig.~\ref{fig:model}). Details  of calculational
results will appear elsewhere.\cite{Lee07} The bridge site $B_2$ in a FHUC is found to be the most stable with an
adsorption energy $E_a$ = $-$2.36 eV/atom favoring over the same site in UHUC of $E_a$ = $-$2.30 eV/atom. When the
second (third) Tl atom is placed at another bridge site around a neighboring restatom, $E_a$ is decreased by 0.008
(0.01) eV/atom indicating attractive interaction between Tl atoms within FHUC. Therefore a nanoclusdter begins to form
only in FHUCs by this attractive interaction between Tl atoms as observed in STM study.\cite{Vitali} On the other hand
when three atoms of Al (Ga) are adsorbed at on-top sites ($T_4$) on Si atoms in the second layer, the most favored for
a single Al (Ga) atom adsorption, $E_a$ is increased by 0.14 (0.01) eV/atom.\cite{Lee-cpc} Because of the repulsive
interaction between the first two Al (Ga) atoms, Al (Ga) nanoclusters are formed in both half unit cells as observed
also in the previous STM studies.\cite{Jia,Chang}

Although the STM image of the surface with Tl nanoclusters best developed at 0.2 ML appears fuzzy,
Pb nanoclusters having about similar atomic mass or In nanoclusters belonging to the same elemental
group exhibit atomically well resolved STM images. One may notice that the Pb and In nanoclusters
are formed with central Si adatoms significantly displaced by adsorbate atoms.\cite{Li2,Li} For Tl
nanoclusters no evidence of such a substrate reconstruction is found in PES or in HREELS data since
no new surface state induced by Tl is seen in PES spectra or any loss peak associated with Tl
indicating a strong chemical bonding has been observed. Zotov et. al., reported, however, that the
Tl adsorbed surface revealed several stable atomic structures depending on $\theta$ when annealed
at a mild temperature, which showed well resolved STM images. Therefore the surface with Tl
nanoclusters at RT is a metastable surface where Tl nanoclusters are thought to be quite mobile to
produce such fuzzy STM images.\cite{Vitali} Such a mobile nature of Tl nanoclusters has also been
suggested by our earlier x-ray study thus proposing a dynamical model where four different atomic
configurations of Tl nanoclusters coexist at RT.\cite{Kim}

\begin{figure}[t]
\includegraphics{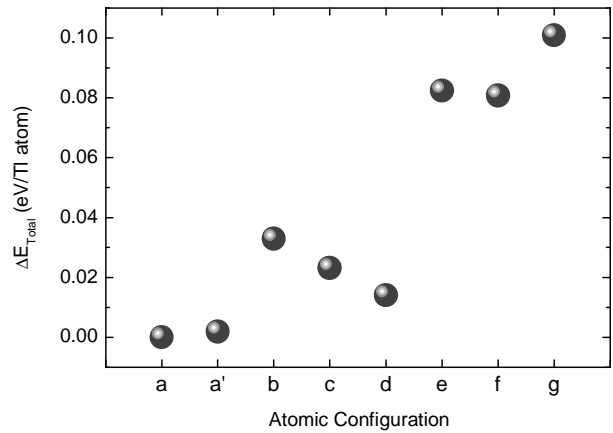}
\caption{\label{fig:energy} Difference in total energy per Tl atom of the eight atomic configurations shown in Fig. 5.
One notes that the maximum total energy difference among the configurations is not greater than 0.10 eV.}
\end{figure}

With this background, we have calculated the hopping rate of Tl atoms to estimate the mobility at
RT for eight different atomic structures depicted in Fig.~\ref{fig:model}. Possible structures
other than those eight models are ruled out since occupation of bridge sites is favored over other
high symmetry sites, for example, the on-top or hollow sites ($H_3$) by 0.07 and 0.10 eV/atom,
respectively. We assumed no reconstruction of substrate surface or replacement of Si atom with
adsorbetes and each Tl nanocluster contains nine Tl atoms as indicated by the coverage of 0.2 ML, 9
Tl atoms on 49 Si atoms, for the fully developed Tl nanoclusters in FHUCs. The total energy
difference calculated for the eight structures is presented in Fig.~\ref{fig:energy}. One finds
that the differences among the structures are less than 0.10 eV/atom. The small total energy
difference among eight possible structures strongly suggests a dynamical model proposed by x-ray
study.\cite{Kim}

In order to estimate the hopping rate of Tl atoms between high symmetry sites, we assumed the half
unit cell of Si(111)-7$\times$7 surface as a potential well with a lateral length of 26.8
\AA~considering the Si dimers as the boundary of the well. This assumption is based on the fact
that the Si dimers constituting the boundary of the unit cell is found to be the most unstable
adsorption sites for metal atoms.\cite{CK}

\begin{figure}[t]
\includegraphics{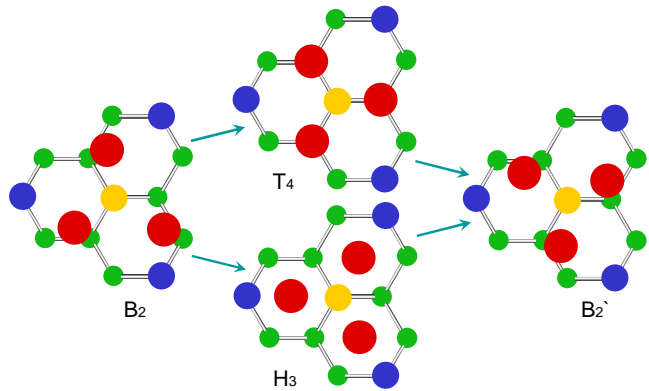}
\caption{\label{fig:hop}(Color online) Two possible hopping paths of Tl atoms from bridge sites
B$_2$ to another bridge sites B$_2^{\prime}$ through (a) $B_2 \rightarrow T_4 \rightarrow B_2'$ and
(b) $B_2 \rightarrow H_3 \rightarrow B_2'$ paths. The energy barriers at on-top site (T$_4$) and at
hollow site (H$_3$) are 0.07 and 0.10 eV, respectively. Tl atoms, Si restatoms, Si adatoms in the
first layer, and Si atoms in the second layer are denoted by red, yellow, blue, and green circles,
respectively.}
\end{figure}

As an example of hopping process, we have considered two processes shown in Fig.~\ref{fig:hop}
where the stable Tl atoms occupying B2 sites initially move to anther stable sites B2' by
overcoming the energy barriers E$_d$=0.07 eV at T4 sites and E$_d$=0.10 eV at H3 sites. To
calculate the hopping rate $\nu$, we adopted the equation $\nu = \nu_0 \rm{exp}(-E_d/k_B T)$, where
$\nu_0$ is attempt frequency, $E_d$ is energy barrier, $k_B$ is Boltzmann constant, and $T$ is
sample temperature. The attempt frequency $\nu_0$ defined as a colliding frequency of atoms to the
energy barrier has been evaluated for Pb, Y, and Ag atoms within the half unit cell as
5$\times$10$^{9}$ hoppings/s.\cite{Vasco} In the calculation by Vasco et. al., they assumed no
hoppings between attractive basins for the convenience of calculation. We note that this value of
$\nu_0$ on the Si(111)-7$\times$7 surface is smaller than typical values on metal surfaces of
10$^{11}$$-$10$^{13}$ hoppings/s.\cite{Vasco} Since $\nu_0$ depends on the width of a potential
well $w$, we have used $\nu_0$=7$\times$10$^{11}$ hoppings/s because of the reduced width $w$=2.2
\AA~between the two bridge sites within a single attractive basin with barriers at $T_4$ or $H_3$
sites. We thus obtain $\nu$=4$\times$10$^{10}$ hoppings/s for the upper path $B_2 \rightarrow T_4
\rightarrow B_2'$ in Fig.~\ref{fig:hop} and 1$\times$10$^{10}$ hopping/s for the lower path $B_2
\rightarrow H_3 \rightarrow B_2'$ at RT. Such significant values of the hopping rate of Tl atoms
clearly indicate that the Tl nanoclusters are quite mobile at RT. Their hopping, however, is
restricted within the same attractive basin of the FHUC because of much higher energy barriers
between the basins and between the neighboring FHUCs.\cite{CK,Wu} Considering the average scanning
frequency of 0.25 scans/s for STM measurement of a half unit cell,\cite{Polop} the fuzzy STM images
of Tl nanoclusters can be caused by such high hopping rates between several metastable atomic
configurations as suggested also by recent x-ray study.\cite{Kim} We thus confirm that Tl
nanoclusters formed on the Si(111)-7$\times$7 surface are quite mobile to rapidly transform their
atomic configurations at RT and show no strong bonding between Tl and substrate Si atoms. The Tl
nanoclusters, therefore, provides an example of self-trapping a nanocluster in nature within a
region as small as nanometer scale.

In addition to such a high mobility, the attractive interaction of a FHUC occupying Tl atom for an
additional Tl atom result  between Tl atoms within an

\section{SUMMARY}
We have measured electronic properties of the crystalline array of Tl nanoclusters formed on Si(111)-7$\times$7 surface
at RT. The valence band PES data show no Tl-induced surface state while intrinsic Si surface states S1 and S2 remain
relatively inert until the Tl coverage of 0.21 ML where Tl nanoclusters cover the entire FHUCs of the surface. Such a
behavior of Si surface states is in sharp contrast with nanoclusters of other atomic species such as Na or Li
nanoclusters where they are rapidly quenched at an early stage of adsorption. No band gap is observed and a
characteristic loss peak associated with Tl adsorption in HREELS data shifts towards the higher loss energy side with
gradual decreasing spectral intensity as Tl nanoclusters are formed. All these experimental data are understood in
terms of the filling of dangling bonds stemming from Si adatoms. Inter-clusters interaction also seems to play a role
in driving the additional shift of S1 state. Our theoretical calculation is found to support our explanation based on
the filling of dangling bonds and further suggests that the Tl nanoclusters are quite mobile to transform their atomic
arrangements by hopping through rather shallow energy barriers between high symmetry binding sites with a rate faster
than 10$^{10}$ hopping/s at RT. We thus conclude that Tl atoms form nanoclusters self-trapped in FHUCs and highly
mobile within the FHUCs with several different structural phases at RT. These unique features of Tl nanoclusters and
the absence of strong Tl$-$Si bondings account for all the experimental data not only discussed here but also other
available observations such as the puzzling fuzzy STM images and a dynamical mixing model proposed by x-ray study. The
system of Tl nanoclusters self-trapped in potential wells of nanometer scale i.e., within the FHUCs of the
Si(111)-7$\times$7 surface, at RT may thus be a good candidate to study Rabi oscillation for numbers of quantum dot
arrays or to explore the possibility as quantum qubits for highly dense optical devices.\cite{XLi}

This work was supported by the Korea Research Foundation Grant funded by the Korean Government (MOEHRD, Basic Research
Promotion Fund (KRF-2006-312-C00513). J. S. Kim acknowledges the support from KISTI (Korea Institute of Science and
Technology Information) under [The Strategic Supercomputing Support Program].


\end{document}